# Temporal assortment of cooperators in the spatial prisoner's dilemma

Tim Johnson[*,†,☼]
Oleg Smirnov[‡]

[†] *Atkinson Graduate School of Management, Willamette University, Salem, OR, USA 97301*
[*] *Center for Governance and Public Policy Research, Willamette University, Salem, OR, USA 97301*
[‡] *Department of Political Science, Stony Brook University, Stony Brook, NY, USA 11794*
[☼] Correspondence: TJ, tjohnson@willamette.edu

**ABSTRACT. We study a spatial, one-shot prisoner's dilemma (PD) model in which selection operates on both an organism's behavioral strategy (cooperate or defect) and its choice of when to implement that strategy across a set of discrete time slots. Cooperators evolve to fixation regularly in the model when we add time slots to lattices and small-world networks, and their portion of the population grows, albeit slowly, when organisms interact in a scale-free network. This selection for cooperators occurs across a wide variety of time slots and it does so even when a crucial condition for the evolution of cooperation on graphs is violated—namely, when the ratio of benefits to costs in the PD does not exceed the number of spatially-adjacent organisms (i.e. $b/c \not> k$).**

## Introduction

Cooperation not only occurs at specific locations, but, also, it takes place at particular points in time. Spatially proximate fruit-bearing plants mast simultaneously to defend against predation[1,cf.2]. Pelicans synchronously form semi-circular patterns in unison to herd prey—a strategy that yields greater hunting success than solo pursuit[3]. Humans meet with neighbors on weekend mornings to clean-up litter in their surroundings.

Existing models of social evolution provide considerable insight into the spatial dynamics of these instances of cooperation and recent models have begun to explore how intra-generational time scales play a role in cooperation's evolution. To consider the latter dimension of social life further, we study a population of spatially distributed organisms playing one-shot prisoner's dilemma (PD) games in which selection operates on both an organism's PD strategy and the time at which it implements that strategy. In so doing, our investigation adds to the literature on the spatial PD[4-8], as well as recent efforts[9,10] to model the role of intra-generational time in social evolution.

Specifically, our model builds on the spatial games framework first proposed in Chapter 8 of Axelrod's *The Evolution of Cooperation*[11] and revolutionized by Nowak and May[4]. Consistent with this paradigm[7,12], we ignore cognitively complex strategies that, for instance, identify and cooperate with kin[13-15], prior cooperators[16,17], phenotypic doppelgangers[18-21], or economic equals[22-24]. Instead, we study a population of organisms, positioned on graphs, who adopt zero-intelligence strategies that always cooperate or defect in one-shot PD games. In these games, cooperators produce a benefit, *b*, less cost, *c*, and they lose that benefit when interacting with a defector who free-rides to steal *b* without incurring a cost. Absent other mechanisms, defectors' higher fitness leads its portion of the population to grow, leaving defectors without cooperators to exploit, thus reducing the population's fitness and producing a worse outcome than if all members of the population had cooperated. Nowak and May[4-6] revealed that playing this game on a grid and updating strategies based on neighborhood comparisons yields clusters of cooperators that persist in the population alongside defectors. Subsequent research explored how modeling subtleties influenced the success of cooperation in this framework[25,26] and it extended the framework's reach into the study of heterogeneous networks[25,27,28], including the development of universal rules to characterize the conditions in which cooperation evolves in any spatial structure[29]. We augment such models by considering how the addition of a temporal dimension to spatial PD models influences social evolution. To do so, we create the possibility for organisms at any node of the spatial structure to choose one out of a fixed set of time slots to implement their behavioral strategy; selection then operates on this choice behavior.

Adding these time slots might appear redundant with existing models. If one regards space and time as interchangeable (for instance, interpreting a three-dimensional model as depicting a two-dimensional space with time as the third dimension), then models of the spatial prisoner's dilemma[6] would seem already to have examined the case we consider. Yet such an interpretation requires a researcher to assume that temporally adjacent organisms interact like spatially adjacent organisms. We do not assume so; in our model, only agents in the same time slot—not adjacent time slots—interact. This distinguishes our model from multidimensional, spatial PD models and it formalizes what we believe is a plausible notion: organisms that stand elbow-to-elbow at the same time likely interact, whereas those that wake at sunset don't interact with those who are fast asleep at that same time, even if they are physically proximate.

Our model might also be mistaken with spatial PD models studying movement. Regardless of whether movement occurs because of stochastic drives[30-32], pre-planned patterns of travel[33], tendencies toward cohesive collective movement[34,35], or the attributes of an organism's current[36-39] or prospective location[40-44], movement constitutes a change in spatial location that occurs across time. Thus, two moving agents that interact implicitly do so at the same moment. In such models, however, the instant of social interaction does not vary. Movement occurs, game play transpires, and this game play either occurs at one particular time point or in a repeated fashion such that organisms interact with others at all available time points. Time of behavioral implementation therefore remains homogenous among organisms in extant models involving movement. Alternatively, we allow organisms in the population to vary when, within a generation, they interact. (Time would play a more salient role in a model of movement in which agents move perpetually within a generation, bumping occasionally into other organisms and commencing a PD game. We know of no such model presently, though we propose it in the discussion.)

By allowing varied times for social interaction, our model also differs from the compelling conceptualization of temporal assortment by Aktipis[45], which focuses on the delayed effects of an organism's behavior on its fitness. A deeper connection exists, however, between our model and studies involving strategic timing[10] and temporal networks[9]. Our model resembles research into strategic timing in that we also examine how cooperation is influenced by the choice of when, within a generation, organisms implement social behavior. Yet, unlike research into timing[10], payoffs in our model do not vary based on their relation to particular events (such as the provision of resources from a public good); instead, they vary based upon the strategies adopted by other organisms active in the same spatial location at the same point in time. Treating the choice of time at which to implement PD behavior as a part of an organism's strategy also differentiates our work from research into the effect of temporal networks on cooperation[9]. Research on temporal networks[9] depicts the time of PD activity as exogenous and studies how the addition of multiple, temporal network layers influences cooperation. Our study allows for the endogenous development of such networks by letting selection operate on organisms' choices of when to implement their behavioral strategy of cooperation or defection.

In particular, we conduct a computer simulation in which organisms adopt strategies consisting of a tuple: (i) a choice in PD play (cooperate or defect) and (ii) the time slot at which to implement that behavior. Time slots, $t$, in the model vary from 1 to 10. When $t$=1, the model amounts to a spatial model in which all spatially adjacent organisms interact with each other. For $t > 1$, organisms at the same spatial location might not interact. Organisms only interact with spatially adjacent organisms who are *also* in their same time slot. If two agents in the same time slot both choose to cooperate in the PD, they receive $b-c$, where $b$ ranges from 1 to 10 and $c$ is fixed at 1; mutual defection earns zero, whereas free-riding earns $b$ and exploited cooperators suffer $-c$. After PD play, organisms' payoffs determine their fitness via the Moran Process; an individual from the population is selected for death at random and another type supplants it with a probability proportional to the strategy's earnings in that generation's PD play[46]. We run this simulation for 10,000 generations, repeating the simulation for 7 runs at each parameter setting.

We perform our simulation on three separate spatial structures—a regular lattice, a small-world network, and a scale-free network. Organisms in the simulated population are arrayed across the nodes of those graphs, with the population size, $N$, set at 100, 225, and 400. In the network models, we vary $k$, the number of edges per organisms, from 2 to 4. We study the small-world network model under two re-wiring probabilities, 0.05 and 0.15.

**Results**

In Fig.1, we present a typical simulation run on our simplest spatial structure—a regular lattice—with $b$=2, $N$=225, and $t$ set at 1, 5, and 10, respectively. In the final generation, defectors dominate the population when $t$=1 (Fig.1(a), leftmost panel) as all agents pool into the only time slot (Fig.1(b), leftmost panel). When $t$=5, cooperators come to dominate the population (Fig. 1(a), middle panel) and selection favors behavioral implementation at multiple time slots (Fig.1(b), middle panel). When we increase the number of time slots to its maximal value, $t$=10, cooperators again grow to fixation (Fig.1(a), rightmost panel) and the population evolves to implement that strategy at 5 different time slots (Fig.1(b), rightmost panel).

Fig. 2 displays the growth and decline of strategies for the example runs displayed in Fig. 1. With $t$=1, the proportion of defectors in the population immediately overtakes that of cooperators (Fig. 2, leftmost panels). However, with $t$=5, cooperators resist the growth of defectors as selection favors cooperators at time slots in which they are populous (Fig. 2, center panels). Close inspection of the lower-middle panel of Fig. 2 indicates selection for defectors in a time slot when the respective number of cooperators and defectors in that time slot are modest; however defectors appear to extinguish themselves as they grow more prolific, driving selection for cooperators in time slots where they already thrive. Temporal assortment of cooperators, in sum, drives the population toward higher levels of cooperation.

As in these example runs, results across simulation runs show selection for cooperators when we add time slots to the lattice structure. When $t=1$ (all agents interact at the same time), cooperators reach fixation in 31.4% of all runs, yet the addition of one additional time slot ($t=2$) results in a slight majority of runs (52.9%) reaching fixation of cooperators. The percent of runs resulting in the fixation of cooperators grows with the addition of each time slot, reaching 80% of all runs at $t=5$ and 86.2% of runs at $t=10$. Indeed, cooperation reaches fixation in 77.6% of all runs when $t≥2$.

As evident in Fig. 3, the frequency of the population reaching fixation of cooperators also grows with the value of *b*. Only in one anomalous case do we find cooperation growing to fixation when $b=1$. However, when $b≥2$, cooperation reaches fixation in 81.1% of all runs. Notably, in violation of the rule for the evolution of cooperation on graphs stipulating that the ratio of benefits to costs must exceed the number of neighboring organisms[29], we find that 54.8% of all runs result in the fixation of cooperators when b/c $\not>$ k (i.e. when b≤4, given $c=1$ and $k=4$) and multiple time slots are present (t≥2).

The panels of Fig.3 suggest that the value of *N* also influences the success of cooperation. When $N=225$, 79.0% of runs result in the fixation of cooperators. When $N=400$, 55.9% of runs do so, whereas 84.1% result in the fixation of cooperators when $N=100$. Does this pattern indicate the success of defectors or the slower growth of cooperators? First, when N=225, 85.7% of all runs result in more than three-quarters of the population adopting the cooperator strategy at G=10000. When N=400, 84.4% of all runs end with more than three-quarters of the population adopting the cooperator strategy at G=10000. Second, on average, when N=100, cooperators reach fixation at generation, G=1709; when N=225, the average generation at which cooperators reach fixation is 3386; when N=400, this average generation is 3502. Together, these pieces of evidence imply that cooperation still grows in larger populations, but it reaches fixation at a slower rate, thus leading to a greater frequency of runs in which cooperation cannot spread through the entire population by G=10000, even though a plethora of agents adopt it.

Analysis of selection on behavioral types and time of implementation in a small-world network yields qualitatively similar results (see Fig. 4). Adding time slots to the model and allowing selection to operate on organisms' choices of which time slot to implement behavior increases the frequency of simulation runs resulting in the fixation of cooperators. When $t=1$, 37.2% of all runs end with the fixation of cooperators; that percent increases to 56.7% when $t=2$, grows to 76.9% when $t=5$, and peaks at 83.9% when $t=10$. Likewise, we observe increasing frequency of cooperator fixation as the value of *b* increases; simulation runs ending with only cooperators occur extremely rarely when b=1, but 45.5% of runs end with cooperator fixation when b=2 and 91.2% of runs do when b=10. As evident in Fig. 4, we again see that the population converges more slowly with larger values of *N*. Additionally, varying the probability of rewiring, *Pr*, does little to influence the simulation's results—as a comparison of the left- and right-half of Fig. 3 indicates. When Pr.=0.05, 71.7%% of all runs result in the fixation of cooperators; a comparable 72.6% of all runs result in the fixation of cooperators when Pr=0.15. Similarly, we find little evidence that rates of cooperator fixation vary across values of *k*: for *k*=2, *k*=3, and *k*=4, we observe, respectively, 72.1%, 72.2%, and 72.0% of simulation runs ending with all agents adopting cooperation.

When we distribute agents across a scale-free network, the population converges to fixation—of either cooperators or defectors—in only 17 out of 6300 (0.27%) simulation runs. Despite a failure to reach fixation, the proportion of cooperators generally increases across generations of the model's simulation runs. The distribution of cooperators' final population proportions, across all runs of the model, balances on a median value of 0.889. In 78.2% of all simulation runs of the scale-free network model, the final percent of cooperators exceeds 75% of the population. These statistics show evidence of growth in cooperators' population share across generations of the scale-free network model, but they appear smaller in magnitude than their analogues calculated from data produced in simulations of the lattice and small-world network models. For instance, the distribution of cooperators' final population proportions rests on median values of 1 in data from each of those models. Likewise, in 85.1% of simulation runs involving the lattice model and 87% of runs involving the small-world network model, the population consists of more than 75% cooperators in the simulation's final generation.

Still, as in our other models, we find in the scale-free network model that cooperators' final population share reaches large values more frequently with higher values of *b* and *t*. Fig.5 displays the proportion of simulation runs, for various values of *b* and *t*, that end with more than 75% of the population adopting the cooperator strategy. When *b*=1, only 14.8% of runs end with cooperators consisting of more than 75% of the population; when *b* =5, 83.3% do and that value increases to 87.3% when b=10. Likewise, when *t*=1, 46.8% of runs end with more than 75% of the population consisting of cooperators, whereas 82.7% and 86.8% of runs end with that percent of cooperators when *t*=5 and *t*=10, respectively.

**Discussion and Conclusion**

Our simulations indicate that selection for time of behavioral implementation spurs the evolution of cooperation in the spatial PD. When *t*>1, we find that cooperators grow—often to fixation—across a wide range of parameter settings, save for when *b*=1. Moreover, we observe cooperators growing to fixation on a regular lattice and a small-world network, even when b/c≯k, so long as multiple time slots are available in which to implement PD play (t≥2). Cooperators regularly increase their proportion of the population at the expense of defectors in models of PD play on a scale-free network, but they do so slowly such that few simulation runs result in the fixation of cooperators. Together, these findings yield insight into a new way that time[9,10] influences social evolution.

The models we study also suggest new ways to extend existing models. For instance, models of movement could be extended to include intra-generational encounters that would occur at various times. That is, when organisms move independent of each other, their movement could cause them to collide at different points of time, thus blending temporal and spatial assortment in substantively meaningful ways worthy of investigation.

Also, the framework used in this paper could be extended to consider more nuanced temporal strategies: instead of specifying strategies as a tuple consisting of an agent's choice in the PD and its selection of the single time at which to implement its PD behavior, strategies could consist of the PD choice and a *list* of time slots at which to play the PD. For instance, periodic behavior (implementing behavior in every other time slot, for example), punch-in-punch-out (implementing behavior in multiple adjacent time slots, but not in others), and random schedules (selecting a number of time slots randomly) could be explored. These variants, moreover, could be explored while considering types that might cooperate or defect variously at different time slots. In sum, the present model can be extended in many ways.

In addition to offering new avenues for theoretical analysis, our model provides novel paths for empirical inquiry. Spatial PD models have received little empirical support in studies designed to test their predictions on either regular lattices[47] or more-complex structures[48,49], thus raising questions about the predictive value of spatial PD models[50-52]. Optimism, about the predictive power of these models, however, has reemerged with recent studies showing that the opportunity to break network links and rewire connections[53,54] yields qualitatively similar results to models of the spatial PD, even when rewiring comes at a cost to participants[55]. Yet such behavior has few non-laboratory analogues outside the sphere of digital networks; moreover, to date, the most-likely mechanism for breaking links in the day-to-day world is movement and it has been shown to have no effect in laboratory studies designed to test its effects in the spatial PD[56]. In light of these studies, our results suggest another potential mechanism that empirical researchers might consider: time of behavioral implementation. Laboratory implementation of multiple time slots for participants arrayed on a spatial structure would constitute a simple extension of existing research designs and has the potential to offer a substantively meaningful mechanism for the link-breaking shown to foster cooperation in previous empirical studies.

Despite these promising pathways, researchers should recognize the limits of our study and future ways to scrutinize those limits. For one, some might contend that we simply study a type of green beard mechanism[20,21] here. At the individual level, we would dispute that claim because the agents in our model cannot discriminate between social partners. However, at the level of the population, we observe a dynamic by which the assortment process can extinguish the presence of defectors in a given time and place, thus giving the appearance of partner discrimination. Also, we make the strong claim that temporally adjacent organisms do not interact like spatially adjacent organisms do. This assumption should be relaxed in future research such that being spatially adjacent and in a neighboring time slot might "wake up" an organism that is not implementing behavior in the same time slot.

Considering ways to enhance the verisimilitude of our model will improve insights into how cooperation can emerge from organisms' decisions about the time at which they implement social behavior. For now, we have shown in a simple model that this mechanism enables the evolution of cooperation via the temporal assortment of cooperators.

**Methods and Materials**

The study's models simulated the evolution of a population of *N* organisms arranged spatially on each of three structures: a regular lattice, a small-world network, and a scale-free network. In the regular lattice, individuals shared edges/connections with *k* other organisms. In the small world network, the parameter *k* determined the average number of connections, while in the scale-free network, *k* seeded the distribution of edges such that the proportion P(●) of organisms bridged by *k* edges/links was distributed $k^\gamma$ (i.e. P(*k*)~$k^\gamma$). At each location, organisms chose a time slot, *t*, when they would interact in the PD and this influenced the partners with whom organisms interacted—only organisms sharing an edge/link and choosing the same time slot would interact with each other.

Interactions took the form of play in a one-shot prisoner's dilemma (PD) game. In the PD, organisms could cooperate or defect. Choosing to defect when a partner chose to cooperate resulted in the payoff *b* (a.k.a. free-riding). Joint cooperation earned *b-c*, whereas joint defection earned 0 and suffering free-riding resulted in –*c*. Organisms adopted strategies that formed a tuple consisting of a behavior in the PD (always cooperate or always defect) and a choice of the time slot at which to implement the PD decision.

Payoffs from the implementation of these strategies influenced the rate at which strategies were adopted and the Moran process was used to execute this selective process. That is, after organisms played the PD, payoffs were tallied and one organism was selected at random to die at the end of the current generation; the organism born as that dead organism's replacement adopted one of the model's strategies with probability proportional to each strategy's success. This process repeated for 10,000 generations, which we label one "run" of the simulation. For every combination of parameters in the model, we ran the model for 7 runs.

The study exogenously varied model parameters to understand whether and how strategy evolution changed according to the values those parameters took. We drew *N* from **N**={100, 225, 400}, *t* from **T**={1, 2, …, 10}, and *b* from **B**={1, 2, …, 10} (note that *c*=1 in all simulations). The parameter *k* was fixed at 4 in the regular lattice, but in both the small-world and scale-free network models, we drew *k* from **K**={2, 3, 4}. The small-world network model was studied under two re-wiring probabilities, 0.05 and 0.15. The simulation explored the full parameter space and did so, as mentioned above, in 7 replications (i.e. runs) for each parameter combination.

Simulations were written in Python 3.7.6. Data generated in the simulations were analyzed in R 3.5.3. All computer code used in the simulations and data analysis is provided in the Supplementary Information. Data sets referenced in the R code are publicly available online and can be accessed via execution of the R code.

**Acknowledgements:** TJ and OS dedicate this work to John M. Orbell (1936-2018).

**Author Contributions:** TJ came up with the idea; TJ and OS planned the research; OS programmed the simulations, with TJ validating and double-checking them; TJ performed the data analysis, with OS double-checking it; TJ wrote the first draft of the paper and OS provided substantial revisions and amendments in subsequent drafts.

**(a)**

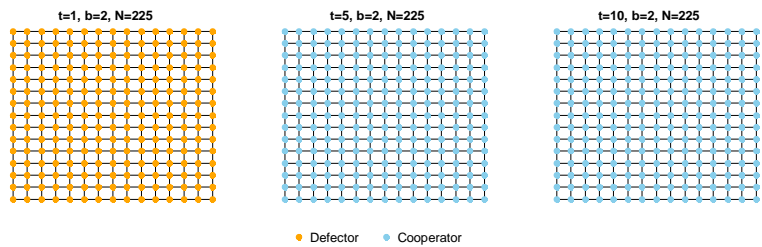

**(b)**

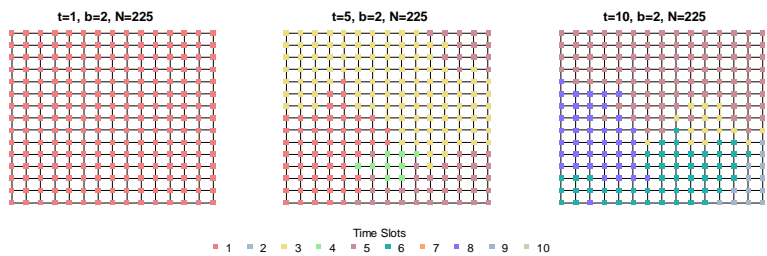

**Fig. 1.** The figure displays the final distribution of behavioral strategies (a) and time slots in which behavior is implemented (b) in an example run of the simulation. The lattices in each column result from data generated in the same simulation run, with parameters set at those listed above each grid. Selection favors defectors when $t=1$ (Fig.1(a), leftmost panel) with all agents pooled into the same time slot at G=10000 (Fig.1(b), leftmost panel). When $t=5$, cooperators come to dominate the population (Fig 1(a), middle panel) and selection leads to behavioral implementation at 4 time slots (Fig.1(b), middle panel). When $t=10$: fixation of cooperators occurs (Fig.1(a), rightmost panel) and the population evolves to implement that strategy at 5 different time slots (Fig.1(b), rightmost panel).

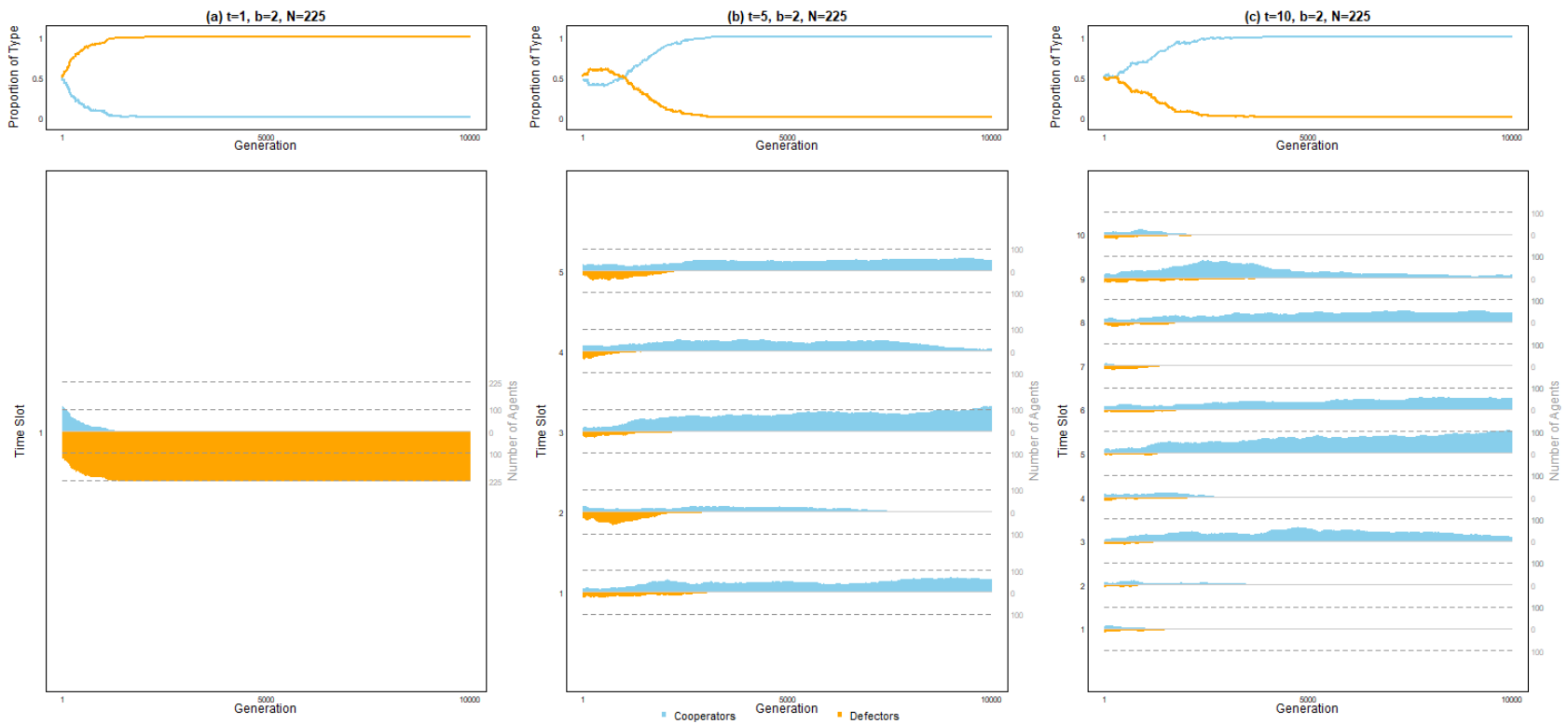

**Fig. 2.** Panels display data from the same example simulation runs displayed in Fig. 1, with parameter settings appearing above the uppermost panel in each column. The upper row of panels displays the respective proportions of cooperators (light blue) and defectors (orange) in the population at each of the 10000 generations of the simulation run. The lower row of panels shows the number of cooperators and defectors in each time slot across generations, with time slots labelled on the left vertical axis and the right vertical axis measuring the number of cooperators and defectors at each time slot. Note that the vertical distance above 0 on the vertical axis measures the number of cooperators and the vertical distance below 0 (i.e. absolute value) measures the number of defectors. When $t$=5 (middle panels) or $t$=10 (right panels), selection favors cooperation and social behavior evolves at multiple time slots.

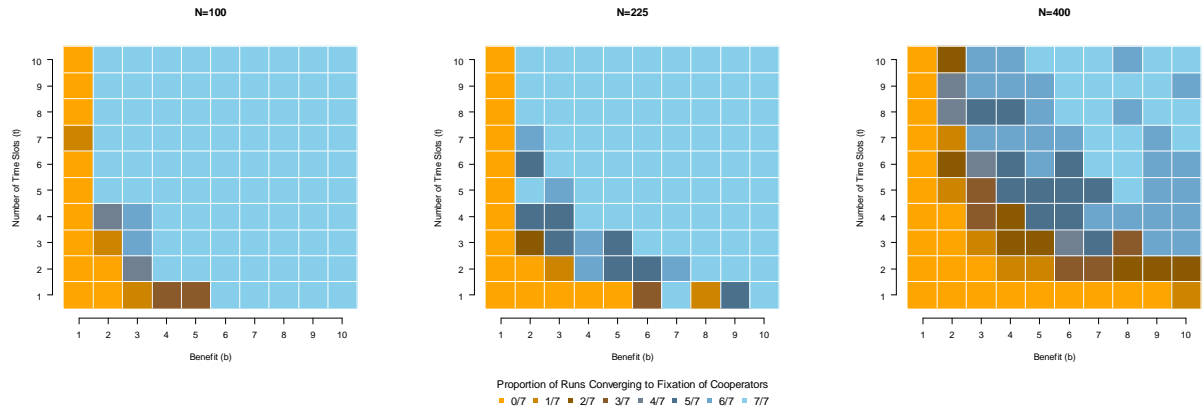

**Fig. 3.** We replicate our simulation for 7 runs at each parameter setting and display how the proportion of runs reaching fixation of cooperators varies by the values of $b$ (horizontal axis) and $t$ (vertical axis), displaying results for each value of $N$ separately (as listed above each panel). Cooperation effectively never reaches fixation when $b$=1, signifying the dearth of gains cooperators obtain even when interacting with each other in such situations. With $b$>2, we observe populations in which cooperators grow to fixation; the prevalence of light blue tiles in the northeast corner of the grids indicates the increased success of cooperation when both $b$ and $t$ take large values. Notably, we observe less instances of cooperators reaching fixation as population size grows; as discussed in the text, this pattern relates to the speed of cooperators' growth as opposed to their ultimate success.

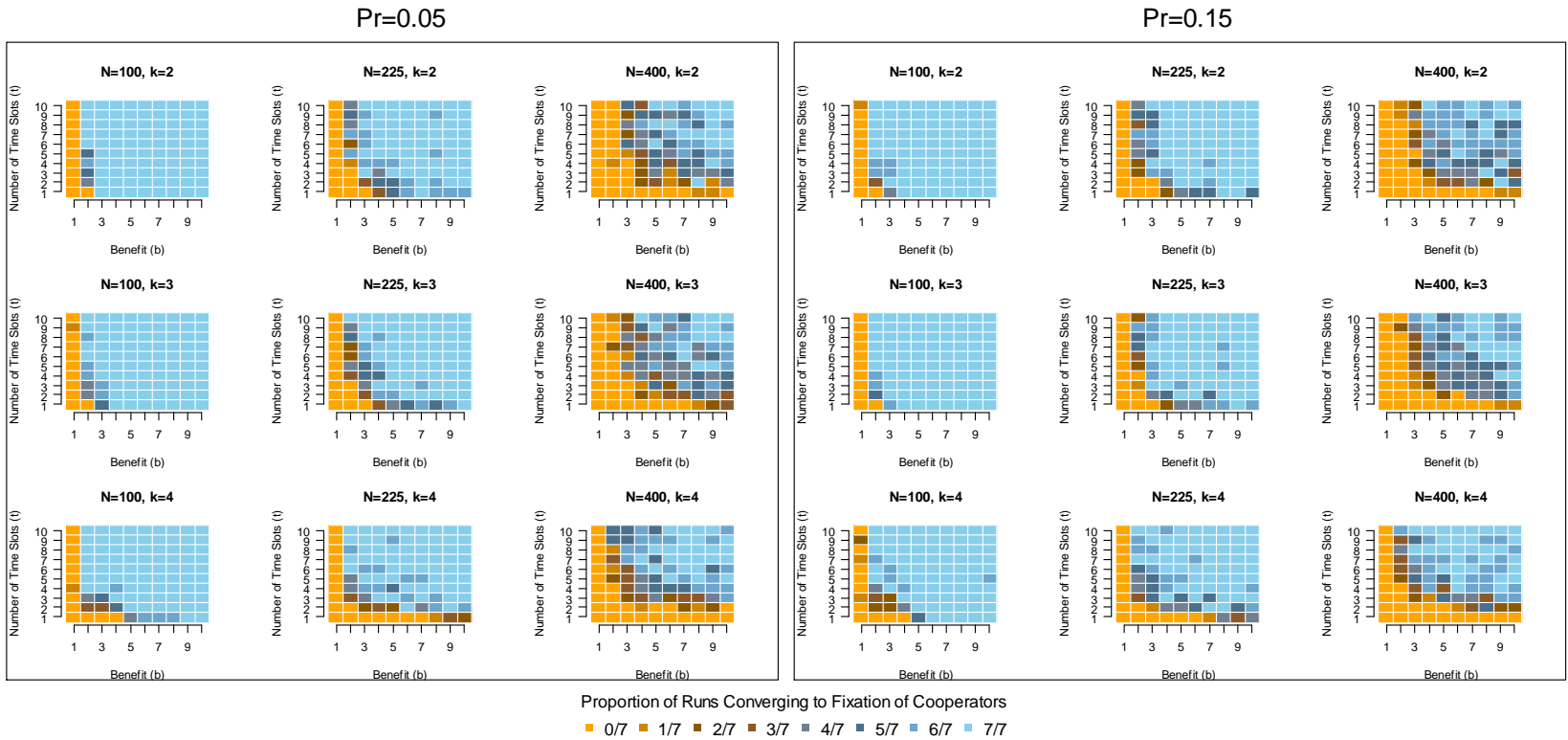

**Fig. 4.** The panels display the frequency of cooperator fixation in simulation runs in which agents reside in a small-world network. Regardless of rewiring probability, *Pr*, we observe cooperator fixation more frequently as *b* and *t* take higher values.

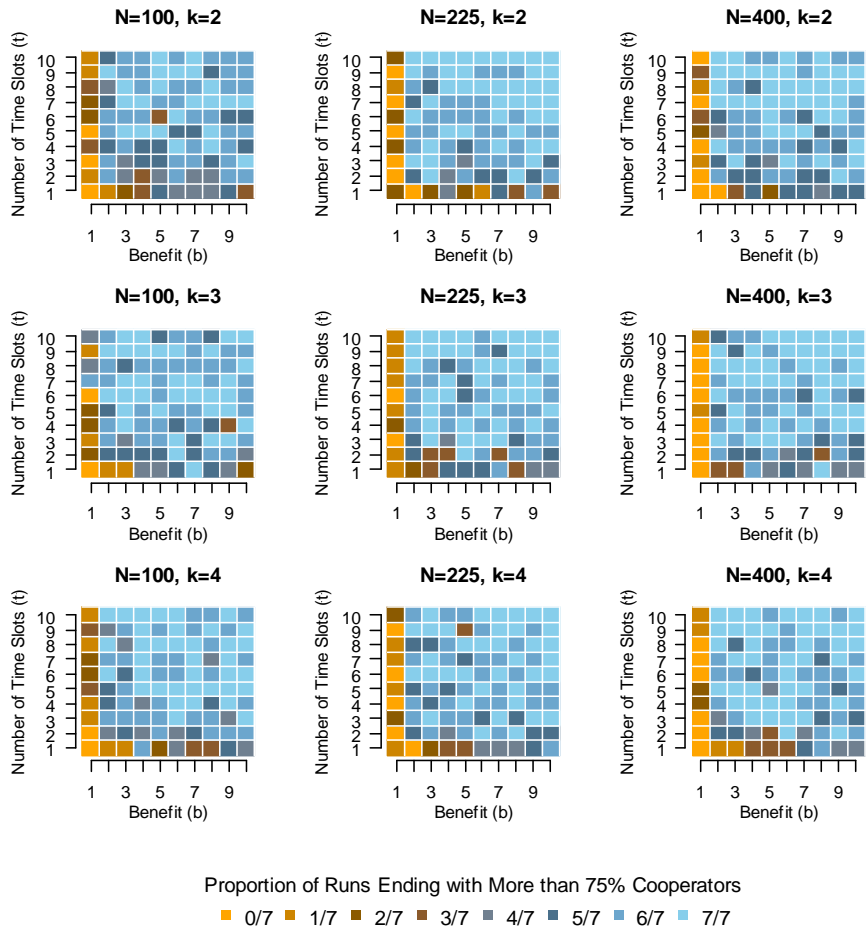

**Fig. 5.** When agents play on a scale-free network, few runs converge to either fixation of cooperators or defectors. We display the proportion of simulation runs at each value of *b* and *t* that end with more than 75% of the population adopting cooperation. Lighter blue tiles indicate a greater frequency of runs, at a given parameter setting, ending with more than 75% cooperators. The results imply slower growth of strategies in the scale-free network, though the general pattern here suggests cooperation occurs more frequently when *t* and *b* take higher values.

**Supplementary Information for**

# Temporal assortment of cooperators in the spatial prisoner's dilemma

SUMMARY. The following hyperlinks provide access to the computer code used in the simulation and data analysis reported in this paper. Clicking on the following links will result in a direct, automatic download of the relevant materials. Data sets used in the analyses are available online and can be accessed by using the R code downloadable in the below links.

A. Python code for [Simulation Models](Simulation Models)

B. R Code for [Figure 1(a)](Figure 1(a)) and [Figure 1(b)](Figure 1(b))

C. R Code for [Figure 2](Figure 2)

D. R Code for [Figure 3](Figure 3)

E. R Code for [Figure 4](Figure 4)

F. R Code for [Figure 5](Figure 5)

G. R Code for [Quantitative Information in the Main Text](Quantitative Information in the Main Text)

16